\documentstyle{article}

\input amssym.def
\input amssym.tex
\def\ch{\mathop{\rm ch}\nolimits}
\def\sh{\mathop{\rm sh}\nolimits}
\begin{document}

\title{Wigner function for free relativistic particles}
\author{O.\,I.~Zavialov, A.\,M.~Malokostov}
\date{}
\maketitle

\begin{abstract}
A generalization of the Wigner function for the case of a free particle with
the ``relativistic'' Hamiltonian $\sqrt{{\bf p}^2+m^2}$ is given.
\end{abstract}

\section{Introduction}

It is well known \cite{WI} that the quantum description
of a  non-relativistic particle, whose momentum space
wave function (at the moment $t=0$) is $\psi ({\bf p})$,
can be handled in the equivalent ``classical'' form. Namely, one defines
the so called Wigner function
\begin{eqnarray}
W({\bf p},{\bf x},t)=\frac{1}{(2\pi)^3}\int d{\bf p}_1\,d{\bf p}_2\,
\psi^*({\bf p}_1)\psi({\bf p}_2)
\delta\biggl({\bf p}-\frac{{\bf p}_1+{\bf p}_2}{2}\biggr) \times
\nonumber
\\
\times
\exp\biggl\{i\biggl(\frac{{\bf p}_1^2}{2m}-\frac{{\bf p}_2^2}{2m} \biggr)t +
i({\bf p}_2 -{\bf p}_1) {\bf x}\biggr\}.
\label{1}
\end{eqnarray}
This quantity can be interpreted (see the  survey \cite{TA})
as the density distribution over the
six-dimensional phase space $\{{\bf p},{\bf x}\}$ of the set of
classical non-relativistic free,
 i.e. corresponding to the Hamiltonian~(\ref{2})
\begin{equation}
H=\frac{{\bf p}^2}{2m}
\label{2}
\end{equation}
particles. One can verify that the expectation values $\overline{f({\bf x})}$
and $\overline{\varphi({\bf p})}$ of the functions $f({\bf x})$
and $\varphi({\bf p})$ over this classical ensemble coincide with the
corresponding quantum matrix elements:
\begin{eqnarray}
\overline{f({\bf x})}&=&\int d{\bf p}\,d{\bf x}\,
W({\bf p},{\bf x},t)f({\bf x})=
\int d{\bf x}\,\widetilde\psi^*({\bf x},t) f(\widehat{\bf x})
\widetilde\psi({\bf x},t),
\label{3}
\\
\overline{\varphi({\bf p})}&=&\int d{\bf p}\,d{\bf x}\,
W({\bf p},{\bf x},t) \varphi({\bf p})=
\int d{\bf p}\,\psi^*({\bf p},t) \varphi(\widehat{\bf p})
\psi({\bf p},t).
\label{4}
\end{eqnarray}

Here
$$
\widetilde\psi({\bf x},t)=\frac{1}{(2\pi)^{3/2}}
\int d{\bf p}\,e^{i{\bf p}{\bf x}}\psi({\bf p},t)
$$
and
$$
\psi({\bf p},t)=e^{-i\frac{p^2}{2m}t}\psi({\bf p})
$$
are the coordinate and the momentum wave
functions at the moment $t$. Moreover,
the expectation value of the function $F({\bf p},{\bf x})$ over this classical
distribution coincides with the matrix element
$$
\langle \widehat F(\widehat{\bf p},\widehat{\bf x})\rangle,
$$
where $\widehat F$ is chosen in Weyl quantization.

So, what most impressive in
the Wigner's idea is that the quantum evolution and the evolution of the
classical ensemble~(\ref{1}) are identical. However, this identity should
not be taken too serious: the Wigner function is not necessarily positive.

It is curious that the generalization of relation (\ref{1}) for the
relativistic case, i.e., for the Hamiltonian
\begin{equation}
H({\bf p})=\omega({\bf p})=\sqrt{{\bf p}^2+m^2},
\label{5}
\end{equation}
seems to be still unknown. The aim of the present paper is to give
such a generalization.

In order to get more insight in what we are going to construct,
let us enumerate the basic properties of the  Wigner function
(1):

a) It is bilinear with respect to $\psi'({\bf p})$.

b) The integral of $W$ over ${\bf x}$ is just the probability density in
momentum space,
$$
\psi'{}^*({\bf p},t)\psi'({\bf p},t)=\int d{\bf x}\,W({\bf p},{\bf x},t).
$$

c) The integral of $W$ over ${\bf p}$ is just the probability density in
coordinate space,
$$
\widetilde{\psi'}^*({\bf x},t)\widetilde{\psi'}({\bf x},t)
=\int d{\bf p}\,W({\bf p},{\bf x},t).
$$

d) The Wigner function admits the classical evolution law. Namely,
$$
W({\bf p},{\bf x},t+\tau)
=W\biggl({\bf p},{\bf x}-\frac{{\bf p}}{m}\tau,t\biggr).
$$
One can easily verify all this but we postpone the corresponding
calculations since these results will follow immediately from the
relativistic formula in the limit of small
momenta.

It is known that the naive quantum mechanical coordinate wave
function
scheme is incompatible
with relativistic principles. Historically this was in fact
the main motivation for quantum field theory.
 For example, one encounters internal difficulties even when
constructing the very coordinate operator $\widehat{\bf x}$~\cite{Wig}.
Anyway, suppose that the free particle with the Hamiltonian~(\ref{5}) has the
momentum space wave function (the function on the upper hyperboloid) equal
to $\psi({\bf p})$. The scalar product between two such functions
$\psi_1({\bf p})$ and $\psi_2({\bf p})$ is
$$
\langle \psi_1({\bf p}),\psi_2({\bf p})\rangle=
\int d\mu({\bf p})\, \psi_1^*({\bf p})\psi_2({\bf p}),
$$
where $d\mu({\bf p})=\frac{d{\bf p}}{\omega({\bf p})}$ is the invariant
measure on the mass hyperboloid. It is generally believed that the
transformation properties of the momentum space wave function
$\psi({\bf p})$ with respect to the Lorentz transformation~$\Lambda$ are as
follows:
$$
\psi_{\Lambda}({\bf p})=\psi({\bf \Lambda p}).
$$
It is natural to go to the momentum space $L_2(d{\bf p},\Bbb R^3)$ of the
wave functions $\psi'({\bf p})$ with the Lebesgue measure $d{\bf p}$. The
corresponding isomorphism is given by the map
$$
\psi'({\bf p})=\frac{\psi({\bf p})}{\sqrt{\omega({\bf p})}}
$$
and the general Lorentz transformation law can be of the form
\begin{equation}
\psi'_{\Lambda}({\bf p})=\sqrt{\frac{\omega({{\bf \Lambda p}})}
{\omega({\bf p})}}e^{i\Omega({\bf \Lambda p})-i\Omega({\bf p})}
\psi'({\bf \Lambda p}).
\label{8}
\end{equation}
Here $\Omega({\bf p})$ is an arbitrary real function.

Thus, the momentum operator $\widehat{\bf p}$ in this latter space is just
the Hermitean multiplica\-tion:
$$
\widehat{\bf p}\psi'({\bf p})={\bf p}\psi'({\bf p}).
$$

It follows from the canonical commutational relations that the coordinate
operators~$\widehat{\bf x}$ (if any) can be chosen as
$\widehat{\bf x}=i\frac{\partial}{\partial{\bf p}}$. Hence, the coordinate
space wave function $\widetilde{\psi'}({\bf x})$ (if any) is again just the
Fourier transform of the momentum space wave function $\psi'({\bf p})$:
\begin{equation}
\widetilde{\psi'}({\bf x},t)=\frac{1}{(2\pi)^{3/2}}
\int d{\bf p}\,e^{i{\bf p}{\bf x}}\psi'({\bf p})
e^{-i\omega({\bf p})t}.
\label{6}
\end{equation}

So, let us formulate our problem.

Find the function $W({\bf p},{\bf x},t)$ which satisfies
the following principles,
inspired by the non-relativistic treatment.

1) It is bilinear with respect to $\psi'({\bf p})$.

2) The integral of $W$ over ${\bf x}$ is just the probability density in
momentum space,
$$
\psi'{}^*({\bf p},t)\psi'({\bf p},t)=\int d{\bf x}\,W({\bf p},{\bf x},t).
$$

3) The integral of $W$ over ${\bf p}$ is just the probability density in
coordinate space,
$$
\widetilde{\psi'}^*({\bf x},t)\widetilde{\psi'}({\bf x},t)
=\int d{\bf p}\,W({\bf p},{\bf x},t).
$$

4) The Wigner function admits the classical
(relativistic) evolution law. Namely,
$$
W({\bf p},{\bf x},t+\tau)
=W\biggl({\bf p},{\bf x}-\frac{{\bf p}}{\omega({\bf p})}\tau,t\biggr).
$$

5) In the limit of small momenta it
tends to the function $W$ defined by~(\ref{1}).

Practically, the requirements 1)--3) are just
the repetition of the corresponding  restric\-tions imposed on the
initial Wigner function (1). Note however the different nature
of the relativistic wave functions ${\psi}$ and ${\psi'}$.

The solution of the problem is given by the formula
\begin{eqnarray}
W({\bf p},{\bf x},t)&=&\frac{1}{(2\pi)^3}\int d{\bf p}_1\,d{\bf p}_2\,
\psi'{}^*({\bf p}_1)\psi'({\bf p}_2)
\delta\bigl({\bf p}-({\bf p}_1\dotplus{\bf p}_2)\bigr) \times
\nonumber
\\
&&\quad
\times
\exp\Bigl(i\bigl(\omega({\bf p}_1)-\omega({\bf p}_2)\bigr)t
+i({\bf p}_2-{\bf p}_1){\bf x}\Bigr).
\label{7}
\end{eqnarray}
Here the symbol $\dotplus$ denotes the special sum on the mass
hyperboloid: if we introduce the four-vectors
$P_1=(\omega({\bf p}_1),{\bf p}_1)$ and
$P_2=(\omega({\bf p}_2),{\bf p}_2)$, then, by definition,
$$
P_1\dotplus P_2\equiv m\frac{P_1+P_2}{\sqrt{(P_1+P_2)^2}},
$$
and ${\bf p}_1\dotplus{\bf p}_2$ is simply the spatial part of
$P_1\dotplus P_2$.

In other words,
\begin{equation}
{\bf p}_1\dotplus{\bf p}_2=m\frac{{\bf p}_1+{\bf p}_2}
{\sqrt{2\bigl(m^2+\omega({\bf p}_1)\omega({\bf p}_2)-
{\bf p}_1{\bf p}_2\bigr)}}.
\label{9}
\end{equation}
The proof of the properties 1)--5) is given in the final section of this
paper.

Of course, the situation with the physical interpretation of the
relativistic function~$W$ is not at all better than of the non-relativistic
one. Namely, this function can be negative as well.  Next, the expectation
values over the classical `` ensemble'' for ``mixed'' functions (depending
simultaneously on ${\bf p}$ and ${\bf x}$) have nothing to do with quantum
matrix elements at least in Weyl quantization.  Moreover, the general
principles of classical mechanics tell us that $W({\bf p},{\bf x},t)$
should be a scalar with respect to Lorentz transformations. This is by no
means consistent with any of the transformation laws of the wave
functions~(\ref{8}).  It shows that even for free particles the
incompatibility of the principles of the conventional quantum mechanics and
 those of special relativity might be even deeper than it is usually
admitted. However, we hope that the function~(\ref{7}) could be still of
use in particular reference frames.  For example, it can be useful in the
theory of ``continuous measure\-ments''. For us it is interesting in order to
analyze the special representation of the canonical commutational relations
we are trying now to apply to quantum field theory~\cite{Mal}.

\section{The proof of the relations 1)--5)}

The property 1) is obvious. The property 5) can be also easily seen
from (9) in
the limit $\omega\rightarrow m$. The property 3) can be verified
immediately. Indeed, the integration of
$W$ over ${\bf p}$ is
absorbed by the $\delta$-function. The remainder is just the
probability density $\psi^*({\bf x},t)\psi({\bf x},t)$. The
procedure leading to the property 2) is also very transparent.
Namely, the integration over ${\bf x}$ produces the $\delta$-function:
$$
\frac{1}{(2\pi)^3} \int d{\bf x}\exp\{i{\bf x}({\bf p}_1-{\bf p}_2)\}
=\delta ({\bf p}_1- {\bf p}_2).
$$
So, effectively
$$
{\bf p}_2={\bf p}_1.
$$
With this relation, the initial $\delta$-function becomes
$\delta ({\bf p}- {\bf p}_1)$ and the integration over ${\bf p}_1$
 leads to the expected equation 2).

Before turning to 4), let us state the following relation. If
${\bf p}=({\bf p}_1\dotplus{\bf p}_2)$, then
$$
\omega({\bf p})=\frac{m}{\sqrt{2}}
\frac{\omega({\bf p}_1)+\omega({\bf p}_2)}
{\sqrt{m^2+\omega({\bf p}_1)\omega({\bf p}_2)-{\bf p}_1 {\bf p}_2}}.
$$
Thus, in order to prove 4), we need only to check that
$$
\omega({\bf p}_2)-\omega({\bf p}_1)  =
({\bf p}_2- {\bf p}_1) \frac{\bf p}{\omega({\bf p})}.
$$
The following chain of transformations makes it obvious.
$$
\omega({\bf p}_2)-\omega({\bf p}_1)  =
\frac{\omega^2({\bf p}_2)-\omega^2({\bf p}_1)}
{\omega({\bf p}_2)+\omega({\bf p}_1)}.
$$
The left-hand side can be rewritten identically as follows.
\begin{eqnarray}
{\omega({\bf p}_2)-\omega({\bf p}_1)} =
\frac{({\bf p}_2-{\bf p}_1)({\bf p}_2+{\bf p}_1)}
{\sqrt{m^2+\omega({\bf p}_1) \omega({\bf p}_2)- {\bf p}_1 {\bf p}_2}}
\times
\nonumber
\\
\times
\frac{\sqrt{m^2+\omega({\bf p}_1) \omega({\bf p}_2)- {\bf p}_1 {\bf p}_2}}
{\omega({\bf p}_1) + \omega({\bf p}_2)}.
\end{eqnarray}
After ovious cancellations we arrive just at the desired relation.

For model considerations, it is convenient sometimes to choose the
two-dimensional space-time. The momentum space corresponding
to the mass-shell $P^2=m^2$ will be one-dimensional:
$$
P=\{m\ch\Gamma, m\sh\Gamma\}.
$$
In this case the $\dotplus$-$\delta$-function can be effectively integrated.
One gets for two dimensions:
\begin{eqnarray}
&&\hskip-2em
\nonumber
W(p,x,t)=\frac{2m^3}{\ch\Gamma}\int_{-\infty}^{+\infty} d\beta\,
\ch(\Gamma+\beta)\ch(\Gamma-\beta)\times
\\
\nonumber
&&\times
{\psi'}^*\bigl(m\sh(\Gamma+\beta)\bigr)
\psi'\bigl(m\sh(\Gamma-\beta)\bigr)
\times
\nonumber
\\
&&
\times
\exp\biggl\{\frac{it}{m}\bigl[\ch(\Gamma+\beta)-\ch(\Gamma-\beta)\bigr]
-\frac{ix}{m}\bigl[\sh(\Gamma+\beta)-\sh(\Gamma-\beta)\bigr]\biggr\}.
\label{11}
\end{eqnarray}

Indeed, we pass to analogous variables for one-dimensional momenta
$p_1$ and $p_2$
in~(\ref{7}):
\begin{eqnarray*}
&&\frac{p_1}{m}=\sh\gamma_1,\qquad\qquad
\frac{p_2}{m}=\sh\gamma_2,
\\
&&\frac{d{\bf p}_1}{m}=\ch\gamma_1\,d\gamma_1,\qquad
\frac{d{\bf p}_2}{m}=\ch\gamma_2\,d\gamma_2.
\end{eqnarray*}

Thus, the Wigner function $W$ takes the form:
\begin{eqnarray}
W&=&m^2\int d\gamma_1\,d\gamma_2\,\ch\gamma_1 \ch\gamma_2\,
{\psi'}^*(m\sh\gamma_1) \psi'(m\sh\gamma_2)\times
\nonumber
\\
&&\quad\times
\exp\biggl\{i\frac{t}{m}(\ch\gamma_1-\ch\gamma_2)
-i\frac{x}{m}(\sh\gamma_1-\sh\gamma_2)\biggr\}\times
\nonumber
\\
&&\quad\times
m\delta\biggl(\sh\Gamma-\frac{\sh\gamma_1+\sh\gamma_2}
{\sqrt{2}\sqrt{1+\ch\gamma_1\ch\gamma_2-\sh\gamma_1\sh\gamma_2}}\biggr).
\label{12}
\end{eqnarray}

The expression under the sign of the square root in the argument of
$\delta$-function can be transformed as follows:
\begin{eqnarray*}
&&
1+\ch\gamma_1\ch\gamma_2-\sh\gamma_1\sh\gamma_2=1+\ch(\gamma_1-\gamma_2)=
\\
&&\qquad=1+\ch^2\Bigl(\frac{\gamma_1-\gamma_2}{2}\Bigr)
+\sh^2\Bigl(\frac{\gamma_1-\gamma_2}{2}\Bigr)
=2\ch^2\Bigl(\frac{\gamma_1-\gamma_2}{2}\Bigr).
\end{eqnarray*}
So, the total integral~(\ref{12}) takes the form
\begin{eqnarray}
W&=&m^3\int d\gamma_1\,d\gamma_2\, \Psi^*(\gamma_1,x,t)\Psi(\gamma_2,x,t)
\delta\Bigl(\sh\Gamma-\sh\frac{\gamma_1+\gamma_2}{2}\Bigr)
\ch\gamma_1\ch\gamma_2=
\nonumber
\\
&=&m^3\int d\gamma_1\,d\gamma_2\, \Psi^*(\gamma_1,x,t)\Psi(\gamma_2,x,t)
\frac{1}{\ch\Gamma}
\delta\Bigl(\Gamma-\frac{\gamma_1+\gamma_2}{2}\Bigr)
\ch\gamma_1\ch\gamma_2,
\end{eqnarray}
where
$$
\Psi(\gamma,x,t)=\psi'(m\sh\gamma)\exp\left(-i\frac{t}{m}\ch\gamma
+i\frac{x}{m}\sh\gamma\right).
$$

Let us go to the new variables $\alpha$ and $\beta$:
$$
\gamma_1=\alpha+\beta,\qquad
\gamma_2=\alpha-\beta.
$$
We arrive finally at (\ref{11}).

The effective ``integration''of the $\delta$-function is also possible
in the three-dimensional case.

This paper was supported by the Russian Foundation for Basic Research,
grant No.~98-01-00162a


\begin{thebibliography}{00}

\bibitem{WI} Wigner E.\,P., Phys. Rev., 1932, {\bf 40}, p. 749.
\bibitem{Wig} Newton T.\,D.,  Wigner E.\,P., Rev. Mod. Phys.,
1949,  {\bf 21}, p. 400.
\bibitem{TA} V.\,I.~Tatarskii, Uspekhi Fiz. Nauk, 1983, {\bf 139},
p.~587 (in Russian).
\bibitem{Mal} O.\,I.~Zavialov, A.\,M.~Malokostov., TMF, to appear.
\end{thebibliography}
\end{document}